# AI-Fuzzy Markup Language with Computational Intelligence for High-School Student Learning

Chang-Shing Lee, *Senior Member, IEEE*, Mei-Hui Wang, Yusuke Nojima, *Senior Member, IEEE*
Marek Reformat, *Senior Member, IEEE,* Leo Guo

*Abstract*—Computational Intelligence (CI), which includes fuzzy logic (FL), neural network (NN), and evolutionary computation (EC), is an imperative branch of artificial intelligence (AI). As a core technology of AI, it plays a vital role in developing intelligent systems, such as games and game engines, neural-based systems including a variety of deep network models, evolutionary-based optimization methods, and advanced cognitive techniques. The 2021 IEEE CIS Summer School on CI for High-School Student Learning was held physically at the JanFuSun Resort Hotel, Taiwan, and virtually on Zoom, on August 10-12, 2021. The main contents of the Summer School were lectures focused on the basics of FL, NN, and EC and the workshop on AIoT (Artificial Intelligence of Things). Invited speakers gave nine courses covering topics like CI real-world applications, fundamentals of FL, and the introduction to NN and EC. The 2021 Summer School was supported by the 2021 IEEE CIS High School Outreach Subcommittee. We also invited students and teachers of high and elementary schools from Taiwan, Japan, and Indonesia. They attended the school and participated in AIoT workshop, gaining experience in applications of AIoT-FML learning tools. According to the short report and feedback from the involved students and teachers, we find out that most participants have quickly understood the principles of CI, FL, NN, and EC. In addition, one of the teachers sent the following remark to the organizers: "This is a great event to introduce students to computational intelligence at a young age, stimulate them to be involved in rapidly evolving fields, and foster participation in future research adventures."

## I. INTRODUCTION

HYBRID learning has become a viral approach for explaining and studying cybernetics systems, especially when the COVID-19 pandemic has swamped the world since 2020. This topic has always been considered as the research domain in the Computational Intelligence (CI) area. CI is traditionally composed of three fields: Neural Networks, Fuzzy Systems, and Evolutionary Computation. CI plays a significant role in developing successful, intelligent systems - from rule-based systems via games to cognitive techniques. Some of the most successful AI systems are based on CI [1].

In 2018 and 2019, the *IEEE CIS Summer School on CI for Human and Robot Co-learning* was held in Taiwan. However, owing to the COVID-19 pandemic, the *2020 IEEE CIS Summer School on CI for Human and Robot Co-learning* was moved to a virtual form using Zoom in Japan and Taiwan. Additionally, the *2020 CI High School Education Program on CI for AI-FML Robotic Learning* supported by IEEE CIS was held in Taiwan as a physical tutorial. Both events aimed at promoting the CI concepts and knowledge to junior high school students and elementary school students in Japan and Taiwan.

The *IEEE CIS High School Outreach Subcommittee*'s goal is to facilitate the outreach to high-school students between the ages of 12-18 and their teachers. In 2021, *IEEE CIS High School Outreach Subcommittee* supported the High School Engagement @ FUZZ-IEEE 2021 [2] and the *2021 IEEE CIS Summer School on CI for High-School Student Learning* [3]. These events were held physically at the JanFuSun Resort Hotel, Taiwan, and virtually on Zoom on August 10-12, 2021 [3]. The goal of the summer school was to gather more high-school students to teach them about the CI knowledge and tools for real-world applications. Figs. 1(a) and 1(b) show the High School Engagement @ FUZZ-IEEE 2021 and opening ceremony @ *2021 IEEE CIS Summer School*, respectively.

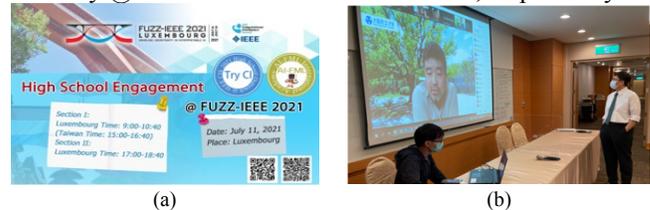

(a)      (b)

Fig. 1. (a) High School Engagement @ FUZZ-IEEE 2021; (b) Opening ceremony @ 2021 IEEE CIS Summer School.

## II. AI-FML WITH IEEE 1855 CO-LEARNING MODEL

### A. AI-Fuzzy Markup Language (FML) and Co-Learning Model

Combining AI with human body language can enhance the genuineness and understanding of human-robot interactions. The robot is not just a tool or device. It becomes a reliable assistant and a friend with whom users can share their thoughts

This work was supported in part by the Taiwan Ministry of Science and Technology under Grant MOST 110-2622-E-024-003-CC1.

Chang-Shing Lee and Mei-Hui Wang are with the Computer Science and Information Engineering Department, National University of Tainan, Taiwan (e-mail: leecs@mail.nutn.edu.tw and mh.alice.wang@gmail.com).

Yusuke Nojima is with the Computer Science and Intelligent Systems Department, Osaka Prefecture University, Japan (e-mail: nojima@cs.osakafu-u.ac.jp).

Marek Reformat is with the Electrical and Computer Engineering Department, University of Alberta, Canada (e-mail: reformat@ualberta.ca).

Leo Guo is with the NUWA Robotics, Taiwan (e-mail: guo.leo@nuwarobotics.com).

and feelings [4].

A Machine-Human Co-learning Model built based on the AI-FML based on IEEE 1855 is illustrated in Fig. 2. The model is suitable for learning different age-group students and is described as follows. The starting point is Human Intelligence (HI). Then, it unfolds into CI machine learning models, including perception abilities based on Neural Networks (NN), cognition skills with Fuzzy Logic (FL), and evaluation methods based on Evolutionary Computation (EC) techniques. During the learning process, the model identifies four stages for young students to gain familiarity with AI-FML: experience-based learning, operation-based learning, practice-based learning, and application-based learning. Finally, the students learn with the support of machines/robots – co-learning – following the integration of elements of CI with real-life applications and learning how to interact with the machines/robots.

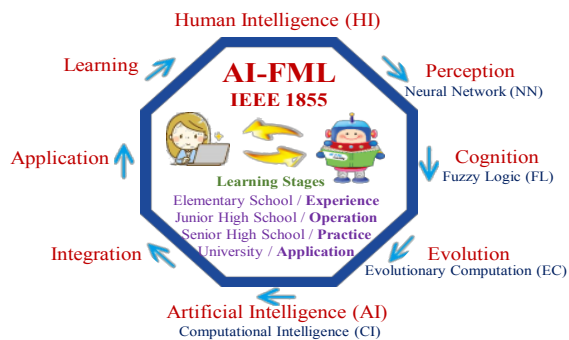
Fig. 2. Machine-human co-learning model based on AI-FML.

### B. Learning Framework of AIoT Workshop based on AI-FML

A learning framework of the AI-FML AIoT (Artificial Intelligence of Things) workshop focused on the elementary-school and high-school students is shown in Fig. 3. It can be described in the following way:

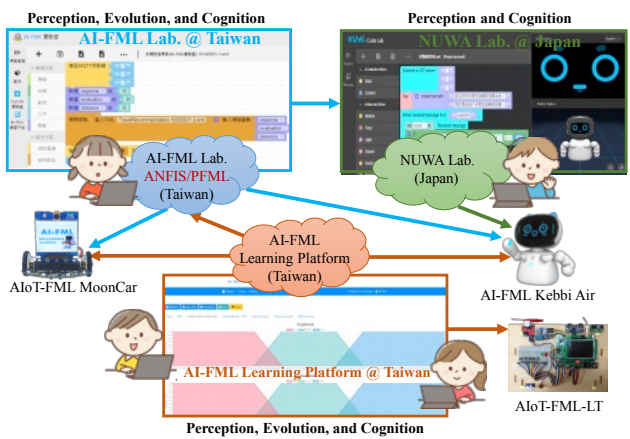
Fig. 3. Learning framework of AIoT workshop based on AI-FML.

1) Students utilize the *AI-FML learning platform* to construct the knowledge base and rule base of the CI-based real-world applications. Additionally, they can apply the Particle Swarm Optimization (PSO) tool to train the knowledge base (KB) and rule base (RB) using the training data set. 2) Students make use of the *AI-FML Lab* to program the AI-FML Blockly. They can operate the fuzzy inference mechanism according to the input data and the learned Adaptive Network-Based Fuzzy Inference System (ANFIS) model, or PSO-based Fuzzy Markup Language (PFML)-based KB and RB [5, 6]. 3) Students adopt the *NUWA Lab* to interact with the *AI-FML robot Kebbi Air*. They can co-learn with the robot to simulate the features of the robot's perception and cognition. 4) The inferred results are sent to the *AI-FML robot Kebbi Air* through the Message Queuing Telemetry Transport (MQTT) server. Besides, the *AI-FML Lab* and *AI-FML learning platform* can directly send inferred results to the *AI-FML robot Kebbi Air* or the *AIoT-FML MoonCar* through the MQTT server and to the *AIoT-FML-LT* through HTTP protocol. 5) Finally, the end-user can communicate with the remote end-user online with such a framework.

### III. HIGH SCHOOL STUDENT LEARNING ON CI

#### A. Information of Participants

The total number of people who registered for *the 2021 IEEE CIS Summer School* for High School Student Learning was 214, including 82, 15, 107, and 10 attendees from Taiwan, Japan, Indonesia, and other countries, respectively. There were 13 graduates, 93 university students, 31 senior high school students, 16 junior high school students, 21 elementary school students, and seven non-students. Figs. 4(a) and 4(b) show the number of participants from different countries and the number of participants from other groups of students, respectively. The percentage of high-school students was around 22%.

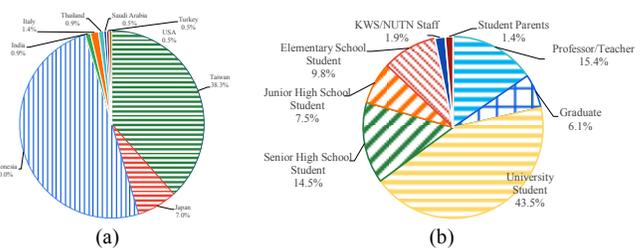
(a) (b)
Fig. 4. (a) Number of participants from different countries; (b) Number of participants from different groups of students.

#### B. Summer School Program for High School Students

Fig. 5 shows the program for the high school students to learn the basics of CI at the Summer School. The main contents of the Summer School are the fundamental pillars of fuzzy systems, neural networks, and evolutionary computation. The lectures were designed to be suitable for beginners who wish to find out more about these areas. The recorded videos of the talks are available as a set of videos, item No. 1 in the Appendix. The lectures introduced the basic concepts of CI for young students or beginners, containing a variety of explanations: (1) capability of fuzzy logic to compute the degree of human perception of, for example, such concepts as hot or cold, especially in the case when different people have different opinions of hot or cold even at the same temperature; (2) the usefulness of neural networks as one of the essential models for machine learning that can compute the mathematical functions; and (3) explanation of evolutionary computation that is based on the observation of the animals' behavior patterns and it is one of the vital machine learning models, too. Integrating emotional intelligence into existing AIoT applications is yet another important aspect of AI [5]. In addition to tutorials on basic concepts of CI, this Summer School provided some lectures on CI-based real-world applications, as well as a QA system with linguistic terms and summarization and a workshop on AIoT to attract high-school students' attention to learning CI.

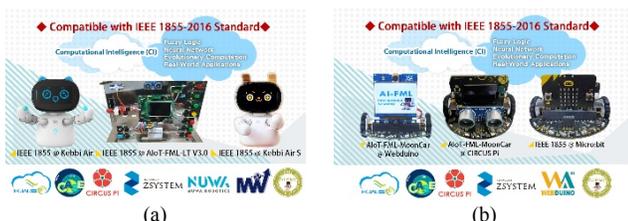

Fig. 5. Program for high school students learning CI at *the Summer School*.

### C. Workshop on AIoT

We have designed and developed the *AI-FML Lab*, *NUWA Lab*, and *AI-FML learning platform* to imitate different aspects of human behavior. In the three AIoT workshops of the *2021 IEEE CIS Summer School for High-School Student Learning*, we focused on the applications to *AIoT-FML intelligent air conditioner* (details are provided in the videos, item No. 2 in the Appendix), *intelligent speaking English*, and *BCI application*, all based on a fuzzy system. The participants were divided into small groups using breakout room at Zoom and each group had a tutor to advise the participants. Fig. 6 shows the AIoT devices including *AI-FML Kebbi Air*, *AIoT-FML-LT*, and *AIoT-FML-MoonCar* that are compatible with IEEE 1855-2016 Standard [7]. Fig. 7(a) shows the AIoT devices at the venue of the *2021 Summer School*. The screen image, the top left of Fig. 7(b), shows a girl tutoring the participants on how to interact with the *AI-FML Kebbi Air* using the AI-FML learning platform during the AIoT workshop.

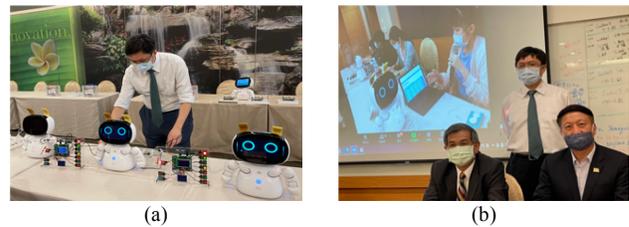

(a) (b)

Fig. 6. (a) IEEE 1855 @ Kebbi Air and AIoT-FML-LT V3.0. (b) AIoT-FML-MoonCar.

There were eight high school student teams and two elementary school student teams. Additionally, when dividing students into groups, we ensured a balanced diversity of students in the context of their origin/country. Each group was asked to perform the following tasks:
- examine the effect of various parameters of fuzzy systems (e.g., the rule base, the number of membership functions) and PSO (e.g., the number of particles and the number of fitness evaluations) on the learning performance.
- design the robot's expressions according to the outputs/results of the fuzzy system.
- present or make a short report on what they learned at the end of the workshop.

The recordings of the participants' co-learning interaction with the machines at the AIoT workshop are available, item No. 3 in the Appendix.

(a) (b)

Fig. 7. (a) AIoT devices; (b) Interaction between the participants and the *AI-FML Kebbi Air*.

### IV. CONCLUSION AND DISCUSSIONS

The *2021 IEEE CIS Summer School for High-School Student Learning* has a significant impact on teaching elements of Computational Intelligence to high-school or university students of computer science, mathematics, electrical engineering, robotics, and related areas. This activity is supported by CIS, which co-funds the Summer School. The scientific goal of the School is to promote CI and expand the attendee's base from college and graduate students to high-school or even elementary school students. Further, such an expansion is in line with the policy of Taiwan and Japan to promote new courses that include concepts of computational thinking as a part of the fundamental education in Taiwan and Japan. Furthermore, it means the involvement of the national and internationally leading researchers in CI, members, and senior members of the IEEE. Therefore, we can say that *The Summer School on Computational Intelligence for High-School Student Learning* is a noteworthy example of teaching the basics of computational intelligence to students from elementary schools, high schools, and universities in Taiwan and Japan. Further, we can say that the AIoT Workshop gives students a good experience in applying the AI-FML Robot and CI-based AIoT-FML Learning Tool for real-world applications.

The total number of individuals who attended at least two-thirds of lectures and the workshops on AIoT was 78, including 48, 9, and 21 from Taiwan, Japan, and Indonesia. All high-school students received the certificate. The ratios of students awarded a certificate of participation was 0.585, 0.600, and 0.196 from Taiwan, Japan, and Indonesia, respectively. The total number of attendees who submitted a short report and feedback survey was 80 and 77, respectively. The submitted reports and surveys showed that participants understood principles of computational intelligence, neural networks, fuzzy logic, and evolutionary computation to a higher degree than in the case of previously organized schools. The targeted audience of this Summer School was high school students. Yet, the registered individuals represented various people from teachers, via graduate students to elementary school students. From our experience and feedback, categorization (e.g., target age) according to the background knowledge is very important for both lecturers and participants to make summer schools fruitful.

We think this event was a well-organized *Summer School*. We were happy to see that the *School* was attended by many participants, both students and teachers, from Taiwan, Japan, Indonesia, and Europe. Finally, we would like to summarize the report citing some feedback comments we received: "*This is a great event to introduce students to computational intelligence at a young age, stimulate them to be involved in rapidly evolving fields, and foster participation in future research adventures.*" And further, "*it is in line with IEEE's mission to advance technology for humanity. Moreover, IEEE offers STEM education opportunities for pre-university students & teachers across the globe. After joining this event, hope they have a big passion & feel motivated to be involved in global issues & trends.*"

ACKNOWLEDGMENT

The authors would like to thank for the support provided by the *IEEE CIS High School Outreach Subcommittee*, the Ministry of Science and Technology (MOST) in Taiwan, JanFuSun Resort Hotel, E. Sun Commercial Bank, Zsystem Technology Company, NUWA Robotics, Tainan City Government, and Kaohsiung City Government in Taiwan.

APPENDIX

TABLE I. SHORT TEXTUAL DESCRIPTIONS OF THE VIDEOS.

| No | Descriptions |
|---|---|
| 1 | **Topic: Slide and Recorded Videos**<br>https://sites.google.com/asap.nutn.edu.tw/2021-ieee-cis-summer-school/lecture-materials<br>**Topic: Onsite Recorded Videos**<br>https://sites.google.com/asap.nutn.edu.tw/2021-ieee-cis-summer-school/videos/recorded-videos<br>**Topic: Fast-Forward Videos**<br>https://sites.google.com/asap.nutn.edu.tw/2021-ieee-cis-summer-school/videos/fast-forward-videos |
| 2 | **Topic:** AIoT-FML intelligent air conditioner application demonstration<br>**Link:** *https://youtu.be/MU_YjneyDbQ*<br>**Descriptions:** Chia-Ying Liu demonstrates AIoT-FML intelligent air conditioner application. |
| 3 | **Topic:** 2021 IEEE CIS Summer School Workshops on AIoT<br>**Link:** *https://youtu.be/Q0-6FYt1394*<br>**Descriptions:** This video shows the participants' co-learning situation with the machines at the AIoT workshop.<br>*Note: This YouTube video provides English subtitles.* |